\documentclass[useAMS,usenatbib,referee]{mn2e}
\usepackage{graphicx}
\usepackage{epstopdf}
\usepackage[figuresright]{rotating}
\usepackage{subfigure,color}
\usepackage{longtable}
\usepackage{xspace}

\newcommand{\ion}[2]{{\textrm{#1}}\,{\textrm{\sc #2}}}

\definecolor{pink}{rgb}{.9,.2,.5}  
\definecolor{purple}{rgb}{.5,.6,.7}

\title[New metallicity calibration for Seyfert 2]
{New metallicity calibration for Seyfert 2 galaxies based on the $N2O2$ index}

\author[Castro et al.]
            { 
               C.~S. Castro$^{1}$,
	      O.~L. Dors$^{1}$\thanks{E-mail:olidors@univap.br}, 
              M.~V. Cardaci$^{2,3}$, 
              G.~F.\ H\"agele$^{2,3}$ \\
$^1$ Universidade do Vale do Para\'iba, Av. Shishima Hifumi, 2911, Cep
12244-000, S\~ao Jos\'e dos Campos, SP, Brazil\\ 
$^2$ Instituto de Astrof\'isica de La Plata (CONICET-UNLP), Argentina. \\
$^3$ Facultad de Ciencias Astron\'omicas y Geof\'{\i}sicas, Universidad Nacional de La Plata, Paseo del Bosque s/n, 1900 La Plata, Argentina.\\
}

\begin{document}

\date{Accepted 0000 Month 00. Received0000 Month 00; in original form 0000 December 17}

\pagerange{\pageref{firstpage}--\pageref{lastpage}} \pubyear{2016}

\maketitle

\label{firstpage}

\begin{abstract}

We derive a new relation between the metallicity of Seyfert 2 Active Galactic Nuclei (AGNs) and the intensity of the
narrow emission-lines ratio $N2O2$=log([\ion{N}{ii}]$\lambda$6584/[\ion{O}{ii}]$\lambda$3727).
The calibration of this relation was performed determining the metallicity ($Z$) of a sample of  58  AGNs through a
diagram containing the observational data and the results of a grid of photoionization models obtained with the {\sc Cloudy} code.
We find the new $Z/Z_\odot$-$N2O2$ relation using the obtained metallicity values and the corresponding
observational emission line intensities for each object of the sample. 
Estimations derived through the use of this new calibration indicate that narrow line regions of Seyfert 2 galaxies exhibit
a large range of metallicities ($0.3 \: \la \: Z/Z_{\odot} \: \la \:2.0$),
with a median value $Z \approx Z_{\odot}$.
Regarding the possible existence of correlations between the luminosity $L(\rm H\beta$),   the electron density,
and the color excess E(B$-$V) with the metallicity  in this kind of objects, we do not find correlations between them.

\end{abstract}


%________________________________________________________________________

\section{Introduction}

Active Galactic Nuclei (AGNs) present in their spectra strong emission lines of heavy elements, easily measured even for objects at high redshifts. Therefore, the 
AGNs metallicities inferred through these emission-lines have an important role in the study of the chemical evolution of galaxies and of the Universe.

The metallicity of the gas phase of AGNs as well as of star-forming
regions can be mainly obtained by two methods. The first one, generally called $T_{\rm e}$-method, is based on measurements of  emission-lines from 
the main ionization stages of a given element (e.g.\ O, N, S) and on direct measurements of the electron temperatures and densities of  the gas \citep{osterbrock06}. 
The second one, called ``strong-line method'', uses a calibration between emission-line ratios easily measured and the metallicity or abundance of a given
element, generally the oxygen (e.g.\ \citealt{pagel79, edmunds84, thaisa98, pilyugin00, pilyugin01, kewley02, dors05, stasinska06, maiolino08, berg11, dors13, brown16, pilyugin16,
valeasari16}).
Concerning the applicability of the $T_{\rm e}$-method, there is a consensus that it  provides reliable metallicity estimations for star forming regions. In fact, 
\citet{pilyugin03} showed that there is a good agreement (at least for the solar neighbourhood) between oxygen determinations based on the $T_{\rm e}$-method
 and the ones derived  through observations of the weak interstellar  \ion{O}{I}$\lambda$1356 line towards the stars (see also \citealt{moss02, deharveng00, 
 rolleston00, meyer98}). However, \citet{dors15} found that the $T_{\rm e}$-method does not work for AGNs. These authors examined the relation between oxygen 
 abundances (generally used
as metallicity tracer) in the narrow-line regions (NLRs) of AGNs estimated from the  $T_{\rm e}$-method,
strong-line method and through central intersect abundances in the host galaxies determined from the radial abundance gradients. They found that the
$T_{\rm e}$-method underestimates the expected  oxygen abundances by until 0.8 dex and that this fact could be due to the
presence of a secondary heating (ionizing) source in addition to the
radiation produced in the inner parts of the AGN \citep[see also][]{zhang13, prieto05, contini12}.
Therefore, the strong-line method seems to be more reliable to be used in AGN metallicity determinations. 

Along decades, several relations between strong emission-lines and oxygen abundances have been proposed  for star-forming regions \citep[see][for a review]{lopez10}. 
Despite metallicities of AGNs have been estimated   by many authors 
(e.g. \citealt{feltre16, dors15, richardson14, batra14, du14, wang11, dhanda07, baldwin03, hamann02,  ferland96, hamann93, hamann92}),
it seems that the unique calibrations available in the literature are  the ones  proposed by \citet{thaisa98} and by \citet{dors14}, who used photoionization model
results to obtain expressions easily applicable considering optical and ultraviolet emission lines, respectively.

With the above in mind, we use a grid of photoionization models to obtain a new relationship between abundances and the strong narrow emission-lines
of Seyfert 2 (Sy2) AGNs observed in the optical spectral range. 
 The present study is organized as follows. In Section~\ref{met} a description 
of the  methodology used to obtain the index calibration is presented. 
In Sect.~\ref{cal} the  calibration obtained is presented.
The discussion and the conclusions of the outcome are given
in Sect.~\ref{resdisc} and Sect.~\ref{conc}, respectively.

\section{Methodology}
\label{met}

To obtain a calibration of the relation between the metallicity  and the strong emission-lines of Sy2 galaxies,  
we compiled  intensities of narrow emission-lines from the literature. These observational data were compared 
with the results of a grid of photoionization models in order to estimate the metallicity of each object. 
In what follows, a description of the photoionization models and of the observational sample are presented.

\subsection{Photoionization models}
\label{mod}

We build a grid of photoionization models using  the version 13.0 of the {\sc Cloudy} code \citep{ferland13}. 
These models are similar to those used by \citet{dors14, dors12}  and the reader is refereed to these works for a detailed description of them. 
In summary, for the Spectral Energy Distribution (SED) we considered  two sources of continuum modelled by: a ``Big Bump'' component 
peaking at 1 Ryd  with a high-energy and 
an infrared exponential cut-off, and a power law with an $\alpha_x = −1$ representing the X-ray source that dominates at high energies taking into account that its 
normalisation must provide an optical to X-ray spectral index $\alpha_{ox}=-1.4$.
This $\alpha_{ox}$ value is the average of the observed values for the entire range of observed luminosities of AGNs by \citet{miller11} and \citet{zamarani81}. Indeed,
photoionization models assuming this SED are able to reproduce optical and infrared observational data of a large sample of AGNs (see \citealt{dors12}).
 In all models, a fixed electron density ($N_{\rm e}$) value of  500 $\rm cm^{-3}$ was assumed. 
It is a representative value for the NLRs densities in AGNs as showed by \citet{dors14}. These authors also showed that Sy2  exhibits electron density values in the range 
$100 \: \la \: \: N_{\rm e} ({\rm cm^{-3}})  \: \la  \: 2000$.
\citet{thaisa98} investigated the influence of the electron density on the  [\ion{N}{ii}]$\lambda$$\lambda$6548,6584/H$\alpha$ 
metallicity indicator, 
finding that  it is suppressed by collisional de-excitation only
for very high density values larger than their critical electron densities,
e.g.\ $N_{\rm e} \approx \: 10^{5} \rm \: cm^{-3}$ (see also \citealt{zhang04}).  
 For the [\ion{O}{ii}] emission lines $\lambda$3726 and $\lambda$3729, the 
critical density is  in order of  2000 and 5000 $\rm cm^{-3}$, respectively. Thus,  effects of electron density 
on metallicity estimations based on [\ion{O}{ii}] lines could be relevant for some
objects studied.

We computed a sequence of models with the logarithm 
of the ionization parameter  ranging from $ -4.0 \: \lid \: \log U \: \lid \: -1.0$, with a step of 0.5 dex.
  $U$  is defined as $U= Q_{{\rm ion}}/4\pi R^{2}_{\rm in} n  c$, where $ Q_{\rm ion}$  
is the number of hydrogen ionizing photons emitted per second
by the ionizing source, $R_{\rm in}$  is  the distance from the ionization source to the inner surface
of the ionized gas cloud (in cm), $n$ is the  particle
density (in $\rm cm^{-3}$), and $c$ is the speed of light \citep{ferguson97, davidson77,shields76, mathews74}.
It is worth mentioning that models with different combination of $Q_{\rm ion}$, $R$ and $n$ 
but that result in the same $U$ are homologous models with
the same predicted emission-line intensities \citep{bresolin99}.

The metallicity range considered was $0.5 \: \lid \:  Z/Z_{\odot} \: \lid \:4.0$. The abundance of all elements was linearly scaled to the solar metal 
composition\footnote{In the {\sc Cloudy} code (version 13.00) the solar oxygen  abundance relative to hydrogen is adopted to be the one  derived by  \citet{alende01}, i.e. 12+log(O/H)=8.69.}, 
with the exception of the N abundance, which was taken from the  following relation between N/O and O/H given by \citet{dopita00}:
  \begin{eqnarray}
       \begin{array}{lll}
 \log({\rm N/H})   & = &  -4.57+\log(Z/Z_{\odot}); \: {\rm for} \:  \log(Z/Z_{\odot}) \lid -0.63,   \\  
    \log({\rm N/H}) & = & -3.94+2\: \log(Z/Z_{\odot});  \: {\rm otherwise}. \\  
     \end{array}
\label{equation:sb2}
\end{eqnarray}

Photoionization model grids assuming these ranges of $U$ and $Z/Z_{\odot}$ values describe emission line intensities observed practically at all wavelengths
\citep{nagao02, nagao06, groves06, dors12, dors14, dors15}.
Models assuming the presence of dust in the gas phase do not reproduce the majority of emission-line intensities of AGNs
\citep{nagao06, matsuoka09, dors14}, hence all the considered models in this work were dust free.

\subsection{Observational sample}
\label{obs}

\begin{figure}
\centering
\includegraphics[angle=-90,width=9cm]{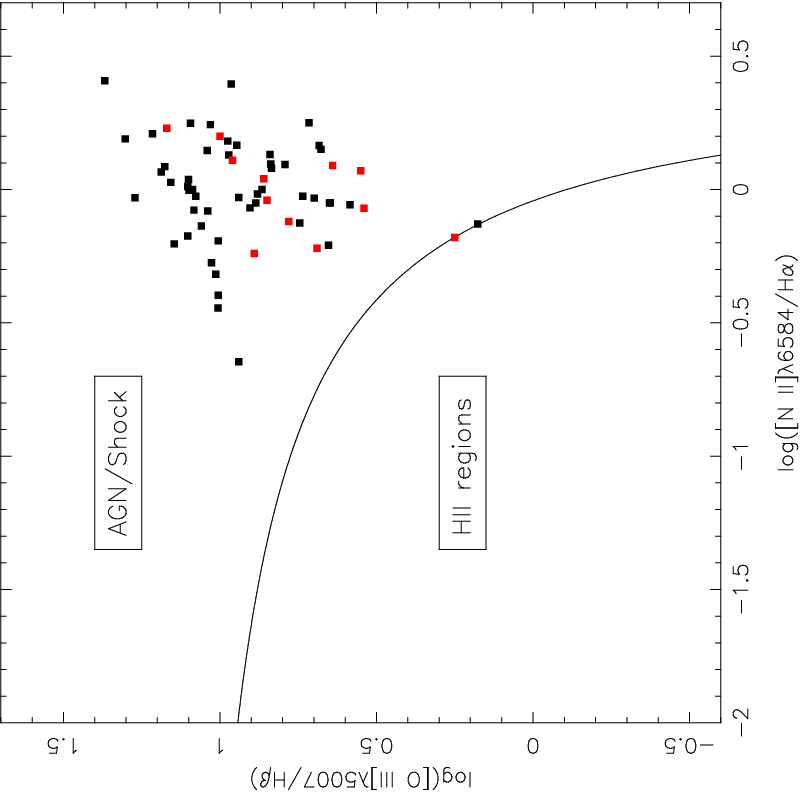}
\caption{$\rm log[O\:III]\lambda5007/H\beta$ vs.\ $\rm log([N\:II]\lambda6584/H\alpha)$ diagnostic diagram.
Solid line, taken from \citet{kewley01}, separates objects ionized by massive stars from those containing active nuclei and/or shock-excited gas
(Equation~\ref{eqk}). 
Black and red squares represent the objects taken from the compilation of  \citet{dors15}  and observed by \citet{dopita15}, 
respectively, listed in Table~\ref{tab1}.}
\label{fdia}
\end{figure}

Intensities of narrow emission lines of AGNs  classified as Seyfert 2 and 1.9 (hereafter Sy2)
observed in the optical range ($\rm 3000 \: \AA  \: < \: \lambda \: < \: 7000 \: \AA$)  were compiled from the literature. 
We did not consider AGNs classified as  Seyfert 1 because these objects  seem to have  shock of gas with high velocity (300-500 km/s;  \citealt{dopita95}),
which is not considered in the {\sc Cloudy} code.
Observational data of LINERs were also not considered because the physics of these objects seems to be little understood and
assumptions of standard photoionization models seem do not reproduce them \citep{thaisa98}.
Our selection criterion was the  measurements of the  intensities of the 
[\ion{O}{ii}]$\lambda$3727,  [\ion{O}{iii}]$\lambda$5007, [\ion{N}{ii}]$\lambda$6584 and [\ion{S}{ii}]$\lambda\lambda$6716,6731
narrow emission-lines.  Observational  data of 47   Sy2  compiled by \citet{dors15} and  13 observed by \citet{dopita15} were considered.
From this sample, only the objects that meet the criteria proposed by \citet{kewley01} to distinguish objects ionized by massive stars from
those containing an active galactic nucleus (AGN) and/or gas shock were considered (see Fig.\ 1). Hence all objects with  
\begin{equation}
\label{eqk}
\rm log[O\:III]\lambda5007/H\beta \: >  \: \frac{0.61}{[log([N\:II]\lambda6584/H\alpha)]-0.47}+1.19
\end{equation}
were selected. 
The final sample consists of 58 objects: 46 compiled by \citet{dors15} and 12 observed by \citet{dopita15}.
In Figure~\ref{fdia} the objects of our  final  sample and a curve representing the  criterion defined by \citep{kewley01}  are shown.
 All objects have redshifts $z<0.1$ and their emission line intensities were reddening corrected.
 Table\,\ref{tab1} lists the identification,   logarithm of the ionization
 parameter,  metallicity, luminosity, electron density, color excess derived along the paper  and the  bibliographic reference for each object of the sample.

\begin{table*}
\caption{Identification,  ionization parameter and metallicity ($Z/Z_{\odot}$) estimated using \textit{interpolation} from Fig.~\ref{f2}, $Z/Z_{\odot}$ through the \textit{N2O2 index}
(Eq.~\ref{eq1}), $\log L$(H$\beta)$, electron density ($N_{\rm e}$), color excess E(B-V) and the original reference for the objects in our sample.}
\label{tab1}
\centering
\begin{tabular}{|lccccccc}	 
\noalign{\smallskip} 
\hline
Identification	  &	$\log U$          &	   \multicolumn{2}{c}{\textit{$Z/Z_{\odot}$ }}        &    $\log L$(H$\beta)$($\rm  erg\: s^{-1}$)  & $N_{\rm e} \: $ ($\rm cm^{-3}$)     & E(B-V)   & Reference \\
\noalign{\smallskip}
\cline{3-4}	 
 \noalign{\smallskip}
                      &                             &   Interpolation   &    \textit{N2O2}   &         &              &          \\
%\noalign{\smallskip}       
\hline	
IZw\,92	           &	      $-$2.5	    &	   0.67		       &         0.67	   &	      41.55						   &	 822.0  			       &    0.15     &  1   	\\
NGC\,3393      &	   $-$2.3	&      1.85	           &          1.80     &	      ---						      &     2022.0				 &   0.20	&  1    \\
Mrk\,176         &	     $-$2.5	   & 	 1.22  		      &       1.12	  &	      40.02						  &	 535.0  			      &   0.60  	&  1  \\
3c033	          &	     $-$2.6	    &	 0.67 		       &      0.66	   &	       40.51						    &	   252.0				 &  0.23	&  1    \\
Mrk\,3	          &	     $-$2.6	    &	  1.25	                &     1.16	   &	       40.91						   &	 948.0  			       &   0.45 	&  1    \\  
Mrk\,573         &	  $-$2.5	  &	1.22	              &       1.12	  &	      40.51						  &	 781.0  			      &  0.30		&  1     \\
NGC\,1068      &       $-$2.4           &     4.00                 &      ---            &           42.03                                                   &      ---                                 &  0.32              &  1 \\
Mrk\,78	          &	    $-$2.7 	   &	  0.32  	        &     0.77	   &	       40.80						  &	 370.0  			      &  0.45		&  1     \\
Mrk\,34	          &	    $-$2.6 	    &	  0.97	               &      0.93	   &	       41.31						   &	  546.0 			       &  0.26  	&  1      \\
Mrk\,1	           &	     $-$2.5	    &	  1.12		        &     1.07	    &		40.20						  &	767.0				      &  0.41		&  1  \\
3c433	          &	    $-$3.0 	    &	 1.02 		        &     1.10	   &		40.36						  &	 50.0				       &  0.57  	&  1   \\
Mrk\,270        &	  $-$2.9	   &	0.87  		       &      0.82	  &	       39.72						 &   1027.0				    &  0.20		&  1  \\
3c452	          &	   $-$3.0  	     &	   1.00	                &     1.02	    &		  40.17 					     &     50.0 				  &  0.47	 &  1   \\
Mrk\,198         &	 $-$2.8    	   &	  1.22		      &       1.16	   &		 40.34  					    &	111.0					&  0.22 	&  1    \\
Mrk\,268        &	  $-$3.0   	    &	  1.17		      &       1.46	   &		40.69						    &	260.0					&  0.40 	&  1   \\
Mrk\,273        &	 $-$3.2    	   &	   0.67	               &      0.62	   &		40.25						   &   50.0					&  0.86 	&  1  \\
NGC\,3227     &	       $-$2.7     	  &	1.62		    &         1.62	 &		 ---							  &   647.0				      &  0.36	&  1	 \\
Mrk\,6	           &	    $-$2.5 	      &	    1.09	        &     1.01	      & 	     ---						     &    647.0 				 &  0.37	&  1   \\
ESO\,138\,G1    &	$-$2.5     	  &	0.57		   &           0.59	  &		 ---							 &    685.0				     &  0.29	&  1	\\
NGC\,5643       &	$-$2.7	     	   &	  0.87	             &         0.82	   &		40.59						    &	141.0				       &  0.52  	&  1   \\
NGC\,1667       &	$-$3.2      	   &	   0.92	              &       0.90	   &		 39.06  					     &  281.0				       &   1.31 	&  1 \\
Mrk\,423       &	  $-$3.2   	     &	     0.82	       &       0.75	     &  	    40.13						  &   239.0				    &	 0.35	&  1	    \\
Mrk\,609        &	   $-$2.7  	      &	      1.50	       &       1.55	      & 	    40.54						 &   239.0				   &  0.57	&  1	\\
Mrk\,226SW     &	$-$3.1     	   &	  0.75		    &          0.69	    &		    --- 						       &    296.0				 &   0.53	&  1       \\
NGC\,3081       &	 $-$2.6	    	    &	   1.32	             &         1.29	     &  	      ---						       &   693.0				  &    0.33	&  1    \\
NGC\,3281       &	 $-$2.7    	    &	    1.32	     &         1.30	     &  	       ---							 &    974.0				   &	0.56	&  1	\\
NGC\,3982       &	$-$2.5     	    &	    1.00	     &          0.95	     &  		---							  &   819.0				    &	 0.24	&  1	\\
NGC\,4388       &       $-$2.5              &      0.92              &         0.83	     &  	       ---							 &   343.0				   &	0.39	&  1	\\
NGC\,5135      &	$-$2.8     	    &	   1.30	              &        1.36	     &  		---							   &   492.0				     &    0.56    &  1     \\
NGC\,5643       &	 $-$2.7    	    &	   0.92 	     &         0.89	     &  	     40.59						   &  451.0				    &	 0.56	    &  1  \\
NGC\,5728       &	 $-$2.7    	    &	   1.30 	     &         1.30	     &  	      41.10						    &  606.0				     &    0.54     &  1   \\
NGC\,6300       &	 $-$2.9    	     &	    0.77	     &         0.73	     &  	     ---							 &    360.0				    &	0.70	&  1	\\
NGC\,6890        &	 $-$2.3    	     &	    1.75	     &     1.61	          &  	      ---							  &    176.0				     &   0.27	&  1	\\
IC\,5063           &	    $-$2.8 	         &     0.87		&    0.82	      & 	       ---							   &	311.0				     &   0.48		&  1  \\
IC\,5135          &	    $-$2.9 	         &	1.07		&    1.09	      & 		 ---							     &    471.0 			       &   0.55 	&  1    \\
Mrk\,744       &	   $-$2.6  	        &	1.62  		&    1.59	      & 	      39.88						     &   606.0  				&    ---	&  1   \\
Mrk\,1066     &           $-$3.0               &       0.95            &    1.01             &                ----                                                    &    ---                                     &    ---         &   1   \\
NGC\,5506      &	 $-$2.7    	      &	     1.19	      &      1.15	   &		   39.68						  &   809.0				     &  0.68		&  1\\
NGC\,2110      &	 $-$3.1    	      &	     1.07	      &      1.12	   &		    39.68						   &   395.0				     &   0.53		&  1\\
NGC\,3281      &	 $-$2.8    	      &	     0.92	      &      0.90	   &		    39.22						    &  471.0				      &   0.68  	&  1\\
Akn\,347	 &	    $-$2.5 	         &	1.37 	         &   1.30	     &  		39.93						       &   606.0				  &    0.54	&  1   \\
UM\,16	        &	   $-$2.6  	         &      0.92		&     0.88	     &  		 41.07  					       &   606.0				  &    0.36	  & 1 \\
Mrk\,533	&	    $-$2.3 	          &	 1.82		 &   1.71	     &  		    40.82						  &   1046.0				     &    0.35    &	1 \\
IZw\,92	         &	     $-$2.5	           &	  0.72		  &  0.68	      & 		      41.44						    &	 805.0  			       &     0.19   &	1  \\
Mrk\,612        &	    $-$2.5 	          &	 1.82 		 &   1.82	     &  		       40.20						    &	  75.0  				&     0.48 &  	 1      \\ 
Mrk\,622        &	   $-$3.2  	         &	 0.57		&    0.54	     &  		      40.71						   &	   64.0 				&     1.17   &	 1      \\    
IC\,1657	 &	    $-$2.5 	          &	  0.85		 &    0.81	      & 		     39.24						     &     40.0 				&    ---		& 2    \\       
IRAS\,01475-0740 &      $-$3.0      	   &	  1.25		  &    1.20       	&			39.80							  &	 90.0				      &     --- 	& 2  \\  
IC\,1816	 &	     $-$2.9	          &	  2.00		 &     2.01		&			39.96							&     8691.0				     &   ---	& 2	   \\  
NGC\,1125     &	     	  $-$2.8               &	 1.00 	      &        0.97	     &  		      39.70						     &     403.0				   &   ---	& 2	\\ 
MCG\,-06-23-038  &	$-$3.0              &	     1.20          &          1.17	  &			 39.43  						&     949.0				      &    ---  	& 2   \\
IRAS\,11215-2806 &	$-$2.8     	    &	  	0.87	   &          0.84	  &			39.22							&      532.0				       &     ---	& 2    \\
ESO\,137-G34    &	 $-$3.0    	      &	  	1.42        &         1.38	   &			  39.74 						  &	 581.0  				&      ---	& 2      \\ 
NGC\,6300	  &	  $-$3.2   	        &	 1.20 	      &       1.17	     &  		   37.92						   &	  330.0 				  &	---	& 2	\\
ESO103-G35	&	$-$2.7     	      &	   	1.25	    &         1.23	   &			39.35						       &     2449.0				    &	    --- 	& 2      \\   
NGC\,6926	  &	  $-$2.8   	         &	 0.90  	      &       0.90	     &  		  38.86 						   &	 305.0  				&	---	& 2	 \\
IC\,1368	     &	     $-$2.7	            &	   0.95	         &     1.02		 &		      38.78						      &     217.0				    &	    --- 	& 2    \\   
NGC\,7590	  &	  $-$2.7   	         &	  0.97	      &        1.00	      & 		  38.54 						    &	 121.0  				 &	 ---	& 2	 \\
\hline 
\end{tabular}
\begin{minipage}[c]{2\columnwidth}
References--- (1) Data compiled by \citet{dors15}. (2) \citet{dopita15}.
\end{minipage}
\end{table*}

The objects compiled by \citet{dors15} were observed using long-slit spectroscopy and those from \citet{dopita15} were observed using integral-field spectroscopy,  therefore,
they constitute an heterogeneous sample, i.e. they were
obtained using different observational techniques and measurement apertures. 
 In Fig.~\ref{fdia} none segregation between the long-slit and integral-field data can be noted. The effects of using such data sample
 do not yield any bias on the abundance estimations (see a complete discussion about this point in \citealt{dors13}).

\section{$Z/Z_{\odot}$-$N2O2$  relation}
\label{cal}

\begin{figure}
\centering
\includegraphics[angle=-90,width=7cm]{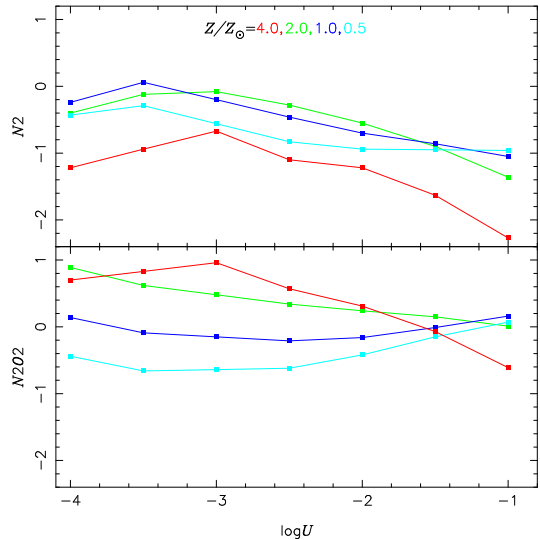}
\caption{$N2O2=\log(\rm [N\:II]\lambda6584/ [O\: II]\lambda3727)$ 
and  $N2=\log(\rm [N\:II]\lambda6584/H\beta)$ versus the logarithm of
the ionization parameter for different metallicities, lower and upper panel respectively. Lines connect
predictions of our models (see Sect.~\ref{mod}) represented by points. Different
colours represents our model results assuming different metallicities, as indicated.}
\label{f1a}
\end{figure}

The $N2O2$ index defined by 
\begin{equation}
N2O2=\rm \log \left( [N\:II]\lambda6584/ [O\: II]\lambda3727 \right)
\end{equation}
was  proposed by \citet{dopita00} to be applied in chemical abundance studies of  star-forming objects (see also \citealt{kewley02, kewley08}).
For star forming regions it  has the advantage to be  little dependent on the ionization parameter than other indexes \citep{kewley02}.
To test if this result is also valid for Sy2s, predictions of our models for the $N2O2$ index and  $N2$=log([\ion{N}{ii}]$\lambda$6584/H$\alpha$)
versus the logarithm of the ionization parameter and for different metallicity values are shown in Fig.~\ref{f1a}.
The $N2$ index was proposed by \citet{thaisa98} to be used as metallicity indicator of AGNs and it 
 has also been applied in studies of star-forming regions (e.g. \citealt{perez09}).
We can see that for each $Z/Z_{\odot} \: \lid \: 2.0$ the $N2O2$ index ranges by about 0.5\,dex for the entire $\log U$ range, while the 
$N2$ ranges by about 1\,dex. For  $Z/Z_{\odot} = 4.0$, a higher variation in both indexes is found. 
 Moreover,  we can note in Fig.~\ref{f1a} (top panel) that the $N2$ indicator does not show a monotonic behaviour with $Z/Z_{\odot}$.

The disadvantage of using strong-line metallicity relation based on the [\ion{N}{ii}] emission lines
is the strong dependence of them on the N/O abundance ratio \citep{perez09}, which is poorly known for AGNs.
 In fact, the  majority of the photoionization model grids  for AGNs has been built considering relations between N and O (or $Z/Z_{\odot}$)
taken from metallicity studies of \ion{H}{ii} regions (e.g. \citealt{dors15, dors14, groves06}). For example, \citet{thaisa98}  assumed in their
models a  secondary  origin for the nitrogen ($\rm N/O\: \sim \: O/H$) and they used a relation  obtained for nuclear starbursts 
derived by  \citet{thaisa94}. However, for the low metallicity regime ($Z\: \la \: 0.3 Z_{\odot}$), the nitrogen seems to have a primary origin
(see e.g. \citealt{pilyugin03a}), which must be considered in AGNs models.

\begin{figure}
\centering
\includegraphics[angle=-90,width=7cm]{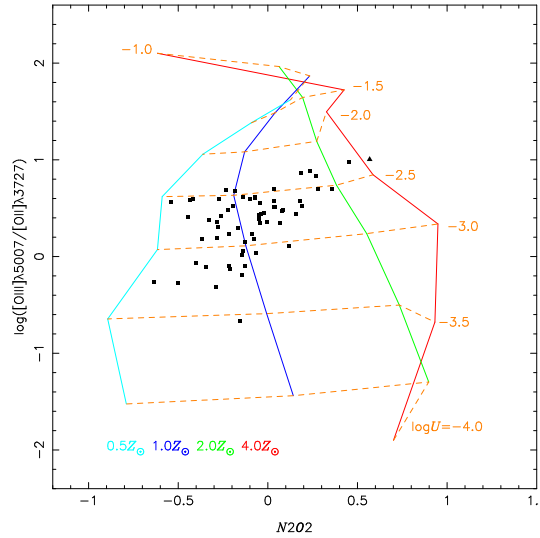}
\caption{log([\ion{O}{iii}]$\lambda$5007/ [\ion{O}{ii}]$\lambda$3727) vs.\ $N2O2$ index.
Solid lines connect our model results (see Sect.~\ref{mod})  
of iso-metallicity, while the dashed lines connect curves of iso-ionization parameter, as indicated.
Points represent  the observational data compiled from the literature (see Sect.~\ref{obs}).
 The point out of the region occupied by the models (represented by a triangle) corresponds to NGC\,1068.}
\label{f2}
\end{figure}

To calibrate the $N2O2$ index with the metallicity, we adopted the following method.
We performed a  [\ion{O}{iii}]$\lambda$5007/[\ion{O}{ii}]$\lambda$3727 versus  [\ion{N}{ii}]$\lambda$6584/[\ion{O}{ii}]$\lambda$3727 diagram
containing the results of our models (see Sect.~\ref{mod}) and the observational data  (see Sect.~\ref{obs}).
In Fig.~\ref{f2} this diagram is shown. We can see that all observational data are located within the regions occupied by our models, 
 with exception of one object, i.e.  NGC\,1068 (represented by a triangle in Fig.~\ref{f2}),   
  not considered in our analysis.

In opposite to the model results of \citet{thaisa98}, it can be seen that the curves representing our photoionization  model results overlap 
for the extreme $\log U$ values. Even though this method can not be used for these extreme values  (i.e. $ \log\:U \: \ga \: -1.8$ and 
$ \log\:U \: \la -3.5$ for $Z/Z_{\odot} \: \ga \:2.0$) since we could not distinguish between two
possible metallicity estimations, we can see from Fig.\,\ref{f2} that our observational data are located in a zone in which models do not overlap between them.
Therefore, to calibrate the metallicity as a function of the $N2O2$ index, we calculated 
 the logarithm of the ionization parameter and the metallicity  values for each object in our sample by linear interpolations between our models.
 These interpolated values are listed in the columns 2 and 3, respectively, of  
Table~\ref{tab1}.

Among the objects selected, we found that NGC\,7674 (not listed in Table~\ref{tab1}) presents the highest measured metallicity value ($Z/Z_\odot$=3.35) in contrast to the relatively low value 
($Z/Z_{\odot} \sim$0.5) found by \citet{dors14} using their C43 index involving near-UV lines. \citet{dors15} also estimated low metallicity values for this 
object  ($Z/Z_{\odot} \sim$1) using the first calibration given by \citet{thaisa98}.  
The optical data of this object was originally published by \citet{shuder81}, who reported the presence of blueward wings on all the emission lines used 
in the present work except on [\ion{O}{ii}]$\lambda$3727. The observed difference in the shape of the emission lines could lead to underestimate the 
[\ion{O}{ii}]$\lambda$3727 flux yielding 
a highest metallicity value.  
Hence, to prevent any possible bias introduced by NGC\,7674 in the calibration we are going to perform, we take off NGC\,7674 from our sample.
The interpolated $Z/Z_{\odot}$ values together with the observational $N2O2$ values was considered and the following equation was obtained
  
\begin{eqnarray}
     \begin{array}{lll}
(Z/Z_{\odot}) \!\!\!& = &\!\!\! 1.08(\pm0.19) \times N2O2^2  +  1.78(\pm0.07) \times N2O2 +1.24(\pm0.01) . \\  
     \end{array}
\label{eq1}
\end{eqnarray}

\begin{figure}
\centering
\includegraphics[angle=-90,width=7cm]{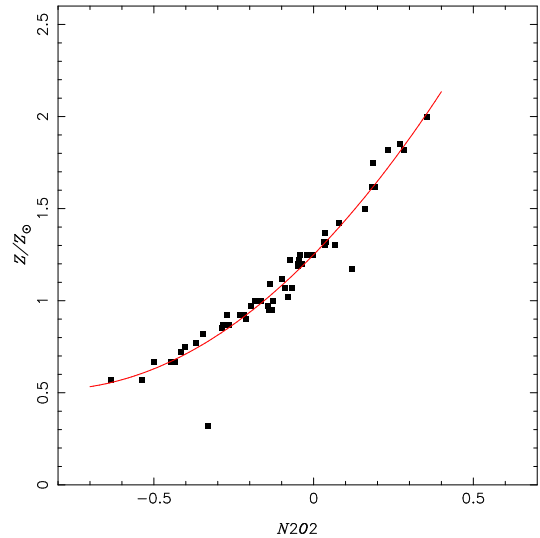}
\caption{$Z/Z_{\odot}$ vs.\ $N2O2$ index.
Points represent metallicity estimations obtained through our photoionization model results (see Sect.~\ref{cal}).
Curve represents the fitting (see Eq.~\ref{eq1}).}
\label{f3}
\end{figure}

In Fig.~\ref{f3} the interpolated $Z/Z_{\odot}$ values as a function of $N2O2$ together with the fitted function are shown.
Although this relation is unidimensional, i.e. it uses only one line ratio mainly dependent on the metallicity, it takes into account 
the dependence with the ionization parameter through the [\ion{O}{iii}]$\lambda$5007/ [\ion{O}{ii}]$\lambda$3727 ratio, which is strongly
dependent on $U$, and that was considered in the $Z/Z_{\odot}$ estimations (see Fig.\,\ref{f2}).

\section{Discussion}
\label{resdisc}

The metallicity  of AGNs is an important parameter because it defines
constraints for the regime of high metallicity in models of chemical evolution of galaxies (e.g. \citealt{cousin16, fu13, pilkington12, molla05}) 
as well as it can  be used to investigate the  enrichment of the Universe (e.g. \citealt{dors14, matsuoka09, nagao06}).

Metallicity estimations of the NLRs of Sy2 galaxies and of the central regions of normal galaxies can be  obtained, indirectly, 
by the use of the central intersect method (e.g. \citealt{dors15, Gusev2012, pilyugin2007, pily04, vanZee1998, Zaritsky1994, VilaCostas1992}).
This method consists in extrapolate to the central regions the radial abundance gradients of  spiral galaxies  estimated from
spectroscopic data of \ion{H}{ii} regions located along their discs. \citet{dors15}
found that abundances obtained by this method are similar or slightly higher than those obtained using strong line methods
for a sample of objects for which there are direct spectral measurements.
However, the central intersect method requires to observe a large sample of
\ion{H}{ii} regions and it is limited to objects spatially resolved, i.e.  objects with low redshifts.
These authors also analysed the results obtained by the use of the $T_{\rm e}$-method
(which involve the weak auroral temperature sensitive emission-lines) finding that this method underestimates the
oxygen abundances by up to $\sim$2 dex compared to the abundances derived through the strong-line method.
Therefore,  determinations based on strong emission-lines of AGNs seem the easiest and most reliable method.

Up to now, it seems that only three relationships between the metallicity or
oxygen abundance and strong and narrow emission-lines of AGNs are available in the literature:
the relationship based on ultraviolet emission lines proposed by \citet{dors14} and two
relations based on optical emission-lines proposed by \citet{thaisa98}.
\citet{dors15} showed that the first relationship proposed by \citet{thaisa98}, given by 
\begin{eqnarray}
       \begin{array}{lll}
{\rm (O/H)}_{{\rm SB98,1}} \!\!\!& = &\!\!\!  8.34  + (0.212 \, x) - (0.012 \,  x^{2}) - (0.002 \,  y)  \\  
         \!\!\!& + &\!\!\! (0.007 \, xy) - (0.002  \, x^{2}y) +(6.52 \times 10^{-4} \, y^{2}) \\  
         \!\!\!& + &\!\!\! (2.27 \times 10^{-4} \, xy^{2}) + (8.87 \times 10^{-5} \, x^{2}y^{2}),   \\
     \end{array}
\label{equation:sb1}
\end{eqnarray}
where $x$ = [N\,{\sc ii}]$\lambda$$\lambda$6548,6584/H$\alpha$ and 
$y$ = [O\,{\sc iii}]$\lambda$$\lambda$4959,5007/H$\beta$, presents a better agreement with recent
photoionization models than the central intersect abundance method. Hereafter, we will compare this relation with our own (Eq.~\ref{eq1}).
 
There are other line ratios that could be used as metallicity indicators of NLRs of Sy2 galaxies, e.g.\ the classical
$R_{23}$ $=([{\rm O\:II}]\lambda3727+[{\rm O\:III}]\lambda4959+\lambda5007)/{\rm H\beta}$
empirical parameter suggested by \citet{pagel79} to estimate oxygen abundances in star-forming regions.
\citet{dors15}, using a grid of photoionization models, found a new O/H-$R_{23}$ relation for NLRs of Sy2 galaxies.
However, these authors pointed out that NLRs of Sy2 galaxies could have a secondary heating (ionizing) source --probably low-velocity shock gas--
in addition to the radiation from the gas falling into the central engine, being the O/H-$R_{23}$ relation not reliable for this kind of objects.
The use of metallicity indicators based on strong emission-lines involving ions with similar ionization potential can minimize this effect.
In this sense, the $N2O2$ index has advantage with respect to the majority  of other metallicity indicators
(see \citealt{kewley02, lopez10, dors11}) because the involved ions,
N$^{+}$ and $\rm O^{+}$, have near ionization potentials, i.e.  29.60 eV and 35.12 eV, respectively.

\begin{figure}
\centering
\includegraphics[angle=-90,width=7cm]{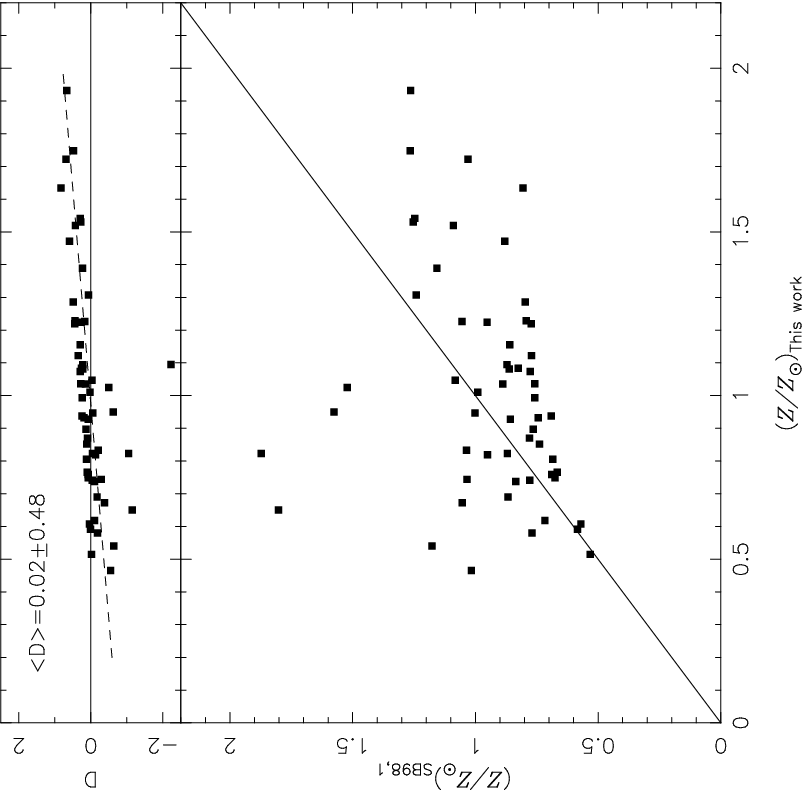}
\caption{Bottom panel: Comparison between $Z/Z_{\odot}$ obtained using the calibrations proposed in this work and the one proposed by \citet{thaisa98} 
for the objects in our sample.
Top panel: Difference  (D) between  metallicity estimations based on our calibration (Eq.~\ref{eq1}) and from the one proposed by \citet{thaisa98}, represented
by Eq.~\ref{equation:sb1}.   The average difference $\rm \langle D \rangle$ between both estimations is indicated.
Dashed line represents a linear regression to these differences, i.e. ${\rm D}=0.89(\pm0.14)\: \times (Z/Z_{\odot})-0.86(\pm0.17)$.}
\label{f4}
\end{figure}

\begin{figure}
\centering
\includegraphics[angle=-90,width=7cm]{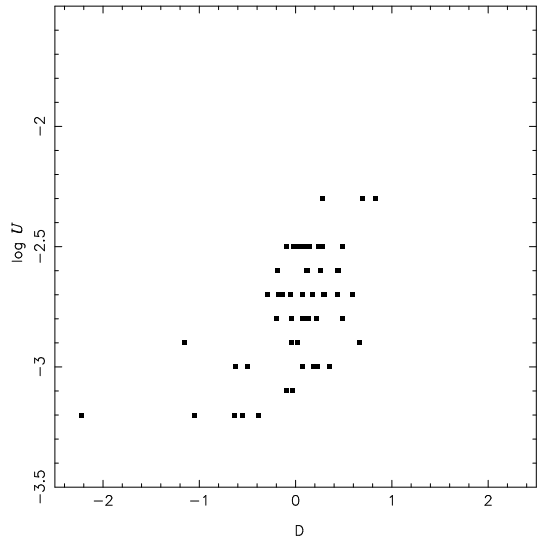}
\caption{Logarithm of the ionization parameter ($U$)  versus the difference  (D) between metallicity estimations 
derived from our calibration (Eq.~\ref{eq1}) and from the one proposed by \citet{thaisa98}, represented
by Eq.~\ref{equation:sb1}.
The $\log U$ and D were taken from Table~\ref{tab1} and Fig.~\ref{f4}, respectively.}
\label{f4r}
\end{figure}

With the aim to compare the  metallicity estimated through our $N2O2$-$Z/Z_{\odot}$ relation with those obtained using the relation proposed by
\citet{thaisa98}, we plotted in Fig.~\ref{f4} both estimations for the objects in our sample  (bottom panel) as well as the difference between them 
 (top panel).  The average difference between both estimations was found to be $\langle D \rangle =0.02 \pm 0.48$ and, based
on the linear regression considering the estimations, it  would seem that the calibration by \citet{thaisa98} yields lower and higher  values for the high 
and low  metallicity regimes, respectively. 
However, the fact that most of the objects are located around ($Z/Z_{\odot}$)=1 could introduce a bias in the calculated linear regression coefficients 
(see upper panel of Fig.~\ref{f4}) due to the low numbers statistics in the extreme metallicity regimes. 
 In order to investigate this discrepancy, the logarithm of the ionization parameter versus the difference between
the estimations is shown in Fig.~\ref{f4r}, where any correlation between can be seen. Therefore, the difference between metallicity estimations 
based on our calibration and those  from the calibration proposed by \citet{thaisa98} probably is due to the use of different N/O-O/H relations
assumed in the models, the evolution of atomic parameters, inclusion of physical process in photoionization model codes 
or due to different methodology considered, hence \citet{thaisa98} considered a theoretical calibration and we a semi-empirical one.

\begin{figure}
\centering
\includegraphics[angle=-90,width=7cm]{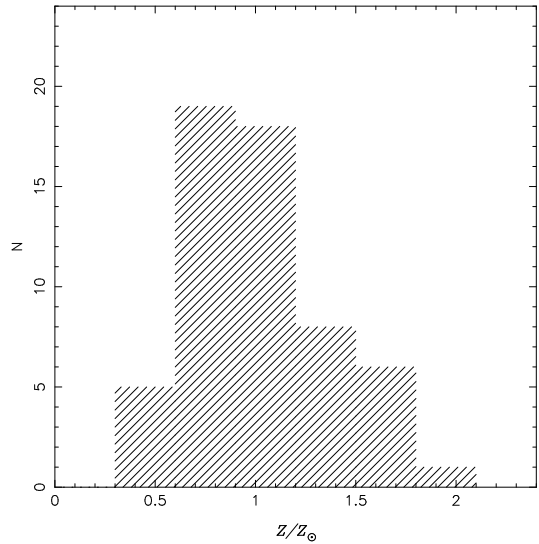}
\caption{Histogram containing the metallicity values derived from the
calibration between $N2O2$-$Z/Z_{\odot}$ (Eq.~\ref{eq1}) for the sample 
of objects listed in Table~\ref{tab1}.}
\label{f4a}
\end{figure}

Analysing the metallicity distribution obtained applying our $N2O2$-$Z/Z_{\odot}$ relation to the objects in our sample (see histogram in Fig.~\ref{f4a}),   we found   
$\sim 55 \%$ of the objects present metallicities in the  $0.75 \: \lid \: Z/Z_{\odot} \: \lid \: 1.25$ range with a median value of $\sim 1.00$  (i.e. 12+log(O/H)=8.69, 
adopting the solar oxygen abundance to be 12 + log(O/H)$_{\odot}$ = 8.69; \citet{alende01}).
The average value of the metallicity considering the whole sample is  $< Z/Z_{\odot} >= 1.03 (\pm 0.38)$, which corresponds to  12+log(O/H)=8.70($\pm 0.13$).
Studying a large sample of star-forming regions, \citet{pilyugin2007, pily06} found, using the P-method \citep{pilyugin01, pilyugin00}, 
that there seems to be a maximum attainable oxygen abundance of 12+log(O/H)$\sim8.87$ ($Z/Z_{\odot} \sim1.50$)  for this kind of regions. 
Most of the objects in our sample show metallicity values lower than this maximum value  derived for star-forming regions. We only  found four   
 objects (NGC\,3393, Mrk\,744, Mrk\,533, IC1816)  with metallicity higher than the maximum estimated by Pilyugin and collaborators. 
\citet{dors15} also calculated the central oxygen abundance for a large sample of Sy2 active objects and star-forming nuclei of normal galaxies founding that most of the 
objects present metallicities in the range $0.6 \: \lid \: Z/Z_{\odot} \: \lid \: 2.0$, with few objects showing higher values.
This is a similar result than the one derived in this work.

\begin{figure}
\centering
\includegraphics[angle=-90,width=7cm]{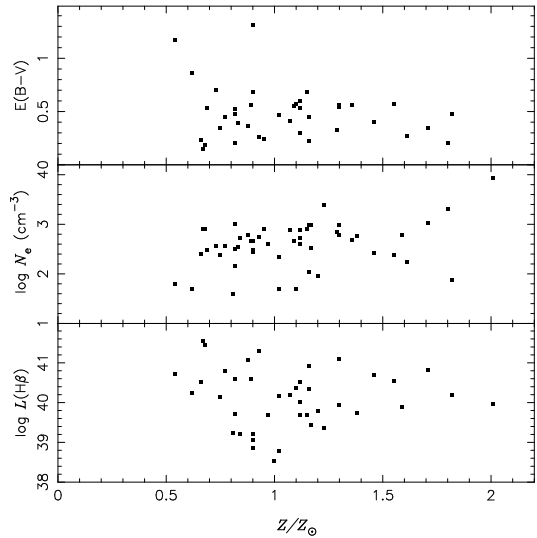}
\caption{log[$L$$(\rm H\beta)$], log[$N_{\rm e}$] and E(B-V) vs.\ $Z/Z_{\odot}$ (bottom, middle and top panel, respectively). 
Points represent estimations for the objects listed in Table~\ref{tab1}.}
\label{f5}
\end{figure}

We investigated whether the metallicity is correlated or not with other Sy2 parameters.
In first place, we analysed the correlation between the AGN luminosity and the metallicity for our sample of Sy2 objects.
 It is well known the existence of a strong correlation between the mass (or luminosity) and
metallicity in ellipticals and spiral bulges (e.g. \citealt{faber73, faber89, Zaritsky1994, lequeux79, 
skillman89, garnett02, pily06}),  in the sense that the most metallic objects exhibit the highest 
mass (or luminosity)  values. This relation seem to be due to action of galactic winds, in which 
massive objects  have deeper gravitational potentials retaining  their gas against the building thermal pressures from supernovae
(see \citealt{hamann99}  and references therein).  
However, it is still  barely known if a similar relation is followed by AGNs.
In fact, \citet{dors14} and \citet{nagao06} found  a slight increase of metallicity, calculated from UV emission lines, 
 with the \ion{He}{ii} luminosity for a large sample of Sy2, Quasar and radio galaxy objects (see also \citealt{hamann93,  hamann99}).
 In order to verify if similar relation is obtained through our estimations, the  luminosity of 46 of our objects were calculated using the published flux of H$\beta$, 
with the redshifts taken from NED\footnote{The NASA/IPAC Extragalactic Database (NED)
is operated by the Jet Propulsion Laboratory, California Institute of Technology, under contract with the National Aeronautics and Space Administration.}
and assuming a spatially flat cosmology with
$H_{0}\,=\,71  \rm \: km\:s^{-1} Mpc^{-1}$, $\Omega_{m}=0.270$,  and $\Omega_{\rm vac}=0.730$ \citep{wright06}.
These values are listed  in the column 5 of  Table~\ref{tab1}.
There are a large scatter in $L({\rm H}\beta$) for each value of $Z/Z_\odot$ (see bottom panel of Fig.~\ref{f5}),
hence, any correlation was found.

 We also  investigated the behaviour of the electron density and of the internal dust content, traced by the color excess E(B-V), as a function  of the metallicity for our sample of objects.
In particular, the existence of  correlations between these parameters  are important 
in building  models of accretion disk  around  black holes (e.g. \citealt{collin99, collin99b})
 The electron density values  (listed   in the column 6  of Table~\ref{tab1})
were estimated from the published
[\ion{S}{ii}]$\lambda\lambda$6716,6731 emission-line intensities and using their relation with the electron density given by
\citet{dors16}.
A large scatter was also found between the estimated densities and the metallicity (see middle panel of
Fig.~\ref{f5}), hence, any correlation was found.
Finally, we analysed the behaviour of the color excess E(B-V) derived from the nebular gas emission lines
as defined by \citet{tucker13} $\rm E(B-V)=1.695\: \log\frac{H\alpha/H\beta}{2.86}$
(listed in the column 7 of Table~\ref{tab1}) as a function of the metallicity (see top panel of  Fig.~\ref{f5}).
As in the previous cases we were not able to find any correlation between these parameters.

\section{CONCLUSIONS}
\label{conc}

We proposed here a metallicity indicator for Sy2 AGNs based on the narrow emission line intensity ratio
$N2O2$=log([\ion{N}{ii}]$\lambda$6584/[\ion{O}{ii}]$\lambda$3727).
The calibration of the relation between $Z/Z_{\odot}$ and the $N2O2$ index was obtained using a sample
of   58 Sy2 galaxies compiled from the literature for which we estimated their metallicities through a  diagram containing the observational 
 log([\ion{O}{iii}]$\lambda$5007/[\ion{O}{ii}]$\lambda$3727,
$N2O2$ ratios and the results of a grid of photoionization models obtained with the  {\sc Cloudy} code.
Using these estimated metallicity values together with the  observational $N2O2$ index values estimated for the objects in our sample,
we calibrated the $Z/Z_{\odot}$-$N2O2$ relation.
Even though this relation depends on only one emission-line ratio, it also depends on the ionization parameter through the
[\ion{O}{iii}]$\lambda$5007/[\ion{O}{ii}]$\lambda$3727 ratio used in the process to calibrate it.
Using the calibration presented in this work, we found that Sy2 galaxies exhibit  a large metallicity range ($0.3 \: \la \: Z/Z_{\odot} \: \la \:2.0$),
with a median value of  $Z \approx Z_{\odot}$.
Hence we did not find any extraordinary chemical enrichment in the narrow line regions of Sy2 AGNs. 
Likewise, any correlation was obtained between metallicity and the H$\beta$ luminosity, the electron density, or 
the color excess E(B$-$V) for the objects in our sample.

\section*{Acknowledgments}
We are grateful to the anonymous referee for his/her useful comments
and suggestions that helped us to substantially clarify and
improve the manuscript.  OLD is grateful to the FAPESP
for support under grant  2016/04728-7 (CADA project).

\label{lastpage}

\end{document}